%% file: main.tex
\documentclass[conference]{IEEEtran}
\IEEEoverridecommandlockouts
\usepackage{cite}
\usepackage{amsmath,amssymb,amsfonts}
\usepackage{algorithmic}
\usepackage{graphicx}
\usepackage{textcomp}
\usepackage{xcolor}
\def\BibTeX{{\rm B\kern-.05em{\sc i\kern-.025em b}\kern-.08em
    T\kern-.1667em\lower.7ex\hbox{E}\kern-.125emX}}

\input{solidity-latex-highlighting.tex}

\begin{document}

\title{Decentralization Paradox: A Study of Hegemonic and Risky ERC-20 Tokens}

\author{\IEEEauthorblockN{Nikolay Ivanov}
\IEEEauthorblockA{\textit{Dept. of Computer Science and Engineering} \\
\textit{Michigan State University}\\
East Lansing, MI, USA \\
ivanovn1@msu.edu}
\and
\IEEEauthorblockN{Qiben Yan}
\IEEEauthorblockA{\textit{Dept. of Computer Science and Engineering} \\
\textit{Michigan State University}\\
East Lansing, MI, USA \\
qyan@msu.edu}
}

\maketitle

\begin{abstract}
In this work, we explore the class of Ethereum smart contracts called the administrated ERC20 tokens. We demonstrate that these contracts are more owner-controlled and less safe than the services they try to disrupt, such as banks and centralized online payment systems. We develop a binary classifier for identification of administrated ERC20 tokens, and conduct extensive data analysis, which reveals that nearly 9 out of 10 ERC20 tokens on Ethereum are administrated, and thereby unsafe to engage with even under the assumption of trust towards their owners. We design and implement \emph{SafelyAdministrated} --- a Solidity abstract class that safeguards users of administrated ERC20 tokens from adversarial attacks or frivolous behavior of the tokens' owners.
\end{abstract}


\section{Background and Definitions}
The role of human in smart contracts is gaining more attention~\cite{ivanov2021targeting}. The person who deploys a smart contract does not receive any privilege or ownership by default. If an Ethereum smart contract needs an owner, administrator, or another privileged user, this functionality must be explicitly implemented by the developer~\cite{antonopoulos2018mastering}. If this governing functionality is not implemented, or if the implemented special role is purely symbolic, such as ownership without any significant power, then we call such smart contracts effectively ungoverned. 
If a smart contract that is not effectively ungoverned, we call it administrated. Another popular smart contract, the TetherUSD stablecoin token, is an example of an administrated token. This token implements a special user role, called owner, which possesses a real privilege that is not available to other users. For example, the owner of the token can create for themselves an arbitrary number of tokens, such as the TetherUSD's \texttt{issue()} function in Fig.~\ref{fig:usdt}. Since the market value of stablecoins is pegged to a fiat currency, United States Dollar in this case --- it would be fair to call the \texttt{issue()} function of this token the owner’s personal dollar printing machine.

\begin{figure}
    \centering
\begin{lstlisting}[language=Solidity]
function issue(uint amount) public onlyOwner {
  require(_totalSupply + amount > _totalSupply);
  require(balances[owner] + amount > balances[owner]);

  balances[owner] += amount;
  _totalSupply += amount;
  Issue(amount);
}
\end{lstlisting}
    \caption{The \texttt{issue()} function of the TetherUSD ERC-20 token.}
    \label{fig:usdt}
\end{figure}

\section{Smart Contracts and ERC-20 Tokens}

\begin{figure}
    \centering
    \includegraphics[width=0.8\linewidth]{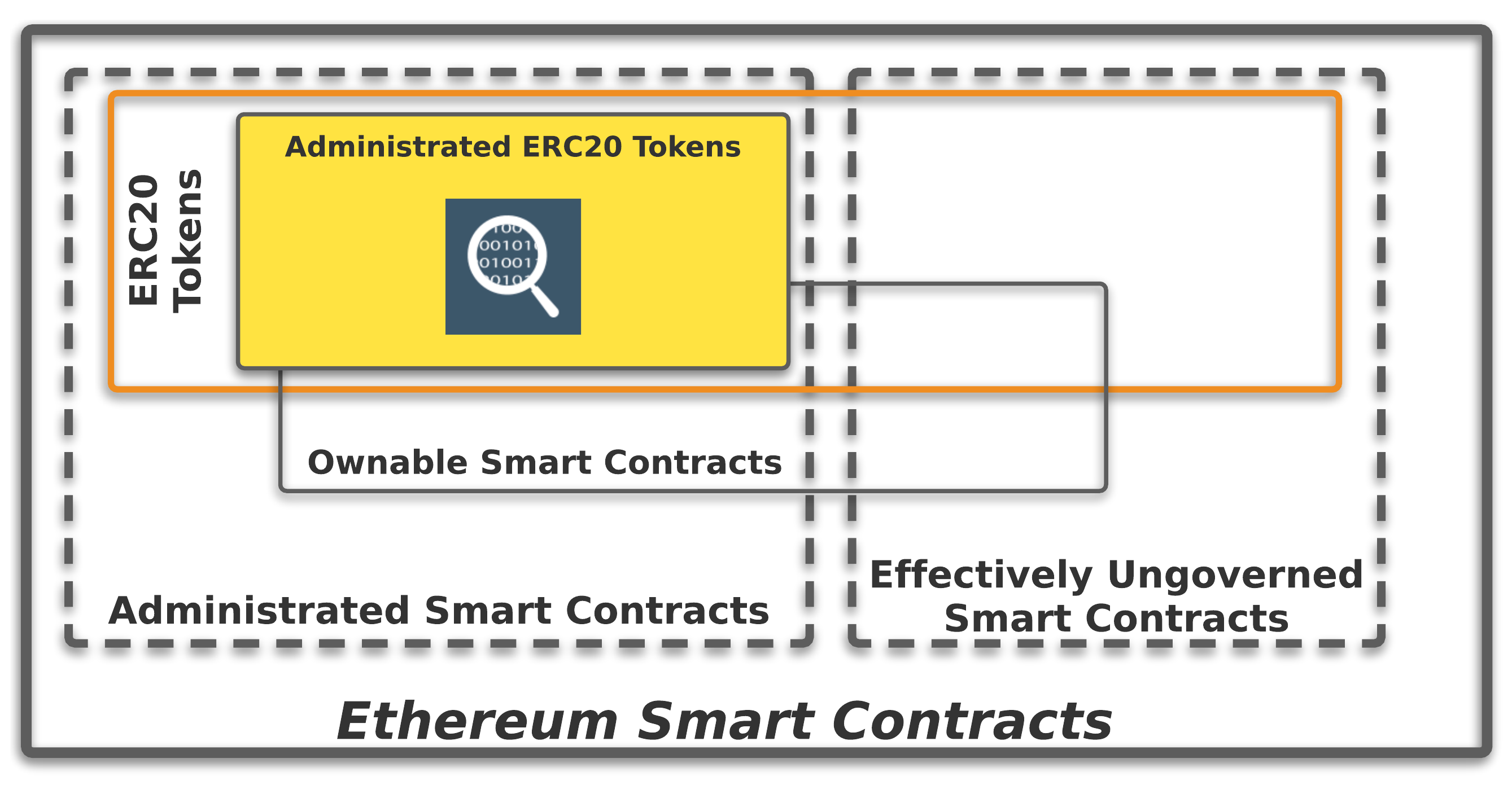}
    \caption{Venn diagram of the types of Ethereum smart contracts.}
    \label{fig:venn}
    \vspace{-5pt}
\end{figure}

Therefore, when it comes to smart contract management, all smart contracts are partitioned into two major groups: the aforementioned administrated smart contracts, and effectively ungoverned smart contracts~\cite{10.1007/978-3-030-86890-1_2} (see Fig.~\ref{fig:venn}). Note that administrated smart contracts is not the same as the popular implementation of ownable smart contracts. An ownership may be purely symbolic, without granting  the owner of the smart contract any impactful privilege. As a result, an ownable smart contract may or may not be administrated. This work focuses on the most popular smart contract type --- fungible ERC-20 tokens, which allow to represent an amountable value, such as digital currencies, gift card balances, reward points, printing quotas, and so on. ERC-20 tokens can be administrated or effectively ungoverned, ownable or not.
In this research, however, we focus on the administrated ERC-20 tokens. Specifically, we pursue the goal of addressing the security issues of Administrated ERC-20 tokens, thereby introduce a new subset of these tokens called Safely Administrated ERC-20 tokens.

\section{Traditional Financial Services versus Administrated ERC-20 Tokens}

Let us juxtapose administrated ERC-20 tokens with traditional financial services. These traditional financial services come in a variety of forms, such as banks, online payment systems, and major credit card providers. Unlike that, ERC-20 tokens are all deployed on the same Ethereum blockchain. The traditional financial services are established businesses. Unlike that, administrated ERC-20 tokens are 100\% online services that can be created by anyone. Traditional services are subject to strict government oversight, while the Administrated ERC-20 tokens are poorly regulated. Traditional financial services often operate upon a hierarchy of executives, managers, and supervisors, which is different from administrated ERC-20 tokens, in which the owner of the privileged private key has an omnipotent power. Even if someone within the structure of the traditional financial services abuses their power, the scope of damage is rarely detrimental. Unfortunately, the administrator of an ERC-20 token can easily steal all the money and disappear. If a traditional financial service is attacked from outside, it normally affects only a small portion of the business, and insurance often covers the damage. With administrated ERC-20 tokens it is not the case: whoever steals the administrator’s private key gets an enormous power. Thus, we can see the following paradox: \textit{the decentralized permissionless nature of Ethereum public blockchain does not automatically guarantee that all the smart contracts deployed on this blockchain are decentralized and permissioned as well}. Specifically, we clearly see that the Administrated ERC-20 tokens exhibit a highly centralized power structure and they are practically less safe than the traditional financial services they are trying to disrupt.

\section{Administrated ERC-20 Patterns}
  \subsection{Self-Destruction}
      One popular administrated pattern is self-destruction, in which a privileged user invokes the \texttt{SELFDESTRUCT} procedure, which permanently blocks any incoming transaction to the smart contract. Most importantly, this procedure transfers any outstanding contract balance to the caller of the function, which is the privileged user.
  
    \begin{figure}
        \centering
\begin{lstlisting}[language=Solidity]
function kill() public onlyAdmin {
  selfdestruct(payable(msg.sender));
}
\end{lstlisting}
        \caption{Example of the self-destruction administrated pattern.}
        \label{fig:self-destruction}
        \vspace{-5pt}
    \end{figure}

  \subsection{Deprecation}
  
Another administrated pattern is smart contract deprecation, as exemplified in Fig.~\ref{fig:deprecation}. Once a smart contract is deployed, its code cannot be modified. The developer cannot directly upgrade a smart contract to address a security vulnerability, fix a bug, or introduce a new feature. To bypass this limitation, the developers of smart contracts implement the functionality in which the privileged user declares the current smart contract as deprecated or obsolete, which forces the current smart contract to redirect all the calls to the respective functions of the upgraded smart contract. Unfortunately, these new functions in the upgraded smart contract can have any code. Therefore, the current deprecation scheme effectively allows the privileged user to inject any arbitrary code in the smart contract.
  
    \begin{figure}
        \centering
\begin{lstlisting}[language=Solidity]
// deprecate current contract in favour of a new one
function deprecate(address _upgradedAddress) public onlyOwner {
  deprecated = true;
  upgradedAddress = _upgradedAddress;
  Deprecate(_upgradedAddress);
}
\end{lstlisting}
        \caption{Example of the deprecation administrated pattern.}
        \label{fig:deprecation}
    \end{figure}

  \subsection{Change of Address}
    \begin{figure}
        \centering
\begin{lstlisting}[language=Solidity]
function setFee(address to) public onlyOwner{
  fee = to;
}
\end{lstlisting}
        \caption{Example of the change of address administrated pattern.}
        \label{fig:change-of-address}
    \end{figure}

In addition to standard Ethereum gas fees, many contracts are designed to charge additional service fees for performing operations with the smart contract. The fees are normally transferred by the smart contract to an address that the privileged user of the smart contract explicitly specifies during the deployment of the smart contract or any time later. However, in our previous work, we demonstrated several attacks that exploit the fee charging functionality and trap clients' funds in the smart contract. For example, if the new fee address is a non-payable smart contract (see Fig~\ref{fig:change-of-address}), the entire transaction involving fee transfer may revert, not allowing users to withdraw their previously deposited funds.

\subsection{Minting and Burning}
Two more common administrated patterns are minting and burning (see Fig.~\ref{fig:minting-burning}), in which the privileged user can create or destroy at their will an arbitrary number of anyone’s tokens. This is by far too much power for a single account.

    \begin{figure}[h]
        \centering
\begin{lstlisting}[language=Solidity]
function mint(address account, uint amount) public onlyOwner {
  _mint(account, amount);
}
function burn(address account, uint amount) public onlyOwner {
  _burn(account, amount);
}
\end{lstlisting}
        \caption{Example of the minting/burning administrated patterns.}
        \label{fig:minting-burning}
    \end{figure}


    




\end{document}

%% file: solidity-latex-highlighting.tex
\usepackage{listings, xcolor}

\definecolor{verylightgray}{rgb}{.97,.97,.97}

\lstdefinelanguage{Solidity}{
	keywords=[1]{anonymous, assembly, assert, balance, break, call, callcode, case, catch, class, constant, continue, constructor, contract, debugger, default, delegatecall, delete, do, else, emit, event, experimental, export, external, false, finally, for, function, gas, if, implements, import, in, indexed, instanceof, interface, internal, is, length, library, log0, log1, log2, log3, log4, memory, modifier, new, payable, pragma, private, protected, public, pure, push, require, return, returns, revert, selfdestruct, send, solidity, storage, struct, suicide, super, switch, then, this, throw, transfer, true, try, typeof, using, value, view, while, with, addmod, ecrecover, keccak256, mulmod, ripemd160, sha256, sha3}, 
	keywordstyle=[1]\color{blue}\bfseries,
	keywords=[2]{address, bool, byte, bytes, bytes1, bytes2, bytes3, bytes4, bytes5, bytes6, bytes7, bytes8, bytes9, bytes10, bytes11, bytes12, bytes13, bytes14, bytes15, bytes16, bytes17, bytes18, bytes19, bytes20, bytes21, bytes22, bytes23, bytes24, bytes25, bytes26, bytes27, bytes28, bytes29, bytes30, bytes31, bytes32, enum, int, int8, int16, int24, int32, int40, int48, int56, int64, int72, int80, int88, int96, int104, int112, int120, int128, int136, int144, int152, int160, int168, int176, int184, int192, int200, int208, int216, int224, int232, int240, int248, int256, mapping, string, uint, uint8, uint16, uint24, uint32, uint40, uint48, uint56, uint64, uint72, uint80, uint88, uint96, uint104, uint112, uint120, uint128, uint136, uint144, uint152, uint160, uint168, uint176, uint184, uint192, uint200, uint208, uint216, uint224, uint232, uint240, uint248, uint256, var, void, ether, finney, szabo, wei, days, hours, minutes, seconds, weeks, years},	
	keywordstyle=[2]\color{teal}\bfseries,
	keywords=[3]{block, blockhash, coinbase, difficulty, gaslimit, number, timestamp, msg, data, gas, sender, sig, value, now, tx, gasprice, origin},	
	keywordstyle=[3]\color{violet}\bfseries,
	identifierstyle=\color{black},
	sensitive=false,
	comment=[l]{//},
	morecomment=[s]{/*}{*/},
	commentstyle=\color{black}\ttfamily,
	stringstyle=\color{red}\ttfamily,
	morestring=[b]',
	morestring=[b]"
}